\documentclass[aps,reprint,pre,showpacs,superscriptaddress]{revtex4-2}
\usepackage{amsmath}
\usepackage{amssymb}
\usepackage{graphicx}
\usepackage{xcolor}
\usepackage{bm}
\usepackage[colorlinks, linkcolor=blue!50!black, urlcolor=blue!50!black, citecolor=blue!50!black]{hyperref}

\newcommand{\reward}{\mathcal{R}}
\newcommand{\treward}{\tilde{\mathcal{R}}}
\newcommand{\mem}{\mathcal{M}}
\newcommand{\Dmut}{\mathcal{D}}
\newcommand{\policy}{\mathcal{P}}
\newcommand{\barv}{\overline{v_0}}
\newcommand{\sgv}{\sigma}
\newcommand{\barth}{\overline{\theta_0}}
\newcommand{\sgth}{\sigma}

\begin{document}

\title{Collective adaptation to time-dependent targets for active Brownian particles}

\author{Gerhard Jung}
\affiliation{Institut f\"ur Theoretische Physik, Universit\"at Innsbruck, 6020 Innsbruck, Austria}
\affiliation{Universit\'e Grenoble Alpes, CNRS, LIPhy, 38000 Grenoble, France}

\author{Eric Bertin}
\affiliation{Universit\'e Grenoble Alpes, CNRS, LIPhy, 38000 Grenoble, France}

\begin{abstract}
We consider a minimal model of active agents which collectively learn to reach a target macroscopic state via a decentralized adaptation process.
Agents are modeled as active Brownian particles which randomly reorient to a given direction through tumbling events.
An average target state is encoded into a reward function evaluated by each agent.
Agents can tune one of the parameters of their microscopic dynamics, called `policy' (e.g., speed or reorientation direction), to optimize their reward function thanks to information exchange with neighboring agents. We describe the evolution of the policy distribution across the agent population with a kinetic theory, under the assumption that the adaptation process is slow with respect to the physical dynamics. The adaptation process is illustrated on several learning protocols with either fixed or time-dependent target velocities. A characteristic learning time $\tau$ combining teaching and mutations emerges from the dynamics. For time-oscillating targets at angular frequency $\omega$, the collective dynamics oscillates in time, with an oscillation amplitude and phase shift determined by the product $\omega \tau$. An unlocking transition to a non-adaptable state is also observed for a rotating target velocity.
\end{abstract}

\maketitle

\section{Introduction}

Active particles, which model for instance active colloids, bacteria or cells, exert forces and produce motion by harvesting energy from their environment \cite{ramaswamy2010mechanics,marchetti2013hydrodynamics,bechinger2016active,ramaswamy2019active}.
Intelligent active particles have recently been introduced as a generalization of the active particle concept, by additionally including internal computation capabilities, taking inspiration from e.g., microrobot experiments \cite{bredeche2022social,ben2023morphological} or crowds of pedestrians \cite{echevarria2023body,bonnemain2023pedestrians}.
Such augmented active particles may be able to autonomously optimize their navigation to a target location through a potentially complex environment \cite{nava2018markovian,liebchen2019optimal,piro2021optimal,piro2022optimal,piro2022efficiency,nasiri2023optimal,mukhopadhyay2026automated,olsen2026information,bilai2026theory},
or more generally to use optimal control \cite{floyd2024learning,cocconi2024dissipation,han2025fluctuation} and reinforcement learning procedures \cite{durve2020learning,monderkamp2022active,boccardo2024reinforcement,munoz2026emergent} to optimize their behavior.
A second important characteristic of intelligent active particles is their ability to communicate and exchange information with other individuals in the population, as in models of animal \cite{durve2020learning,ziepke2022multi} or microorganism behavior \cite{zampetaki2021collective}, or in microrobot experiments \cite{ben2023morphological}.
The joint ability to perform basic computations and to exchange information allows for adaptation processes, which are decentralized when communication is short-ranged 
\cite{ben2023morphological,bredeche2022social,kaspar2021rise,levine2023physics,majidi2019soft,mo2023challenges,cazenille2024signaling,fersula2026aggregating}, giving rise to the emerging field of smart active matter \cite{pishvar2020foundations,cichos2020machine,kaspar2021rise,levine2023physics,goldman2024robot,nasiri2024smart,te2025artificial,baulin2025intelligent,lowen2026towards}.
Minimal models of assemblies of interacting smart agents have been shown numerically to display a rich phenomenology like flocking transitions or autonomous collective decision making \cite{borra2021optimal,yang2022autonomous,negi2025binary,goh2026adaptative,iyer2026emergent}.
Such emergent collective properties naturally call for a statistical description --a paradigm shift from a robotics viewpoint \cite{ben2023morphological,bredeche2022social,janzen2026active,fersula2026aggregating}-- and may require the development of specific coarse-graining methodologies \cite{ziepke2022multi,vansaders2023informational,cocconi2024dissipation,jung2025kinetic,ariosto2025replication,garnier2025hydrodynamics,jung2026theory}.

In this paper, we aim for a statistical description of the collective learning (or adaptation) process occurring in an assembly of simple agents interacting via pairwise information exchange, naturally leading to a kinetic theory type of description \cite{de2011modeling, coscia2011mathematical, burini2016collective}. The generic kinetic theory formalism for such a collective and decentralized learning process has recently been introduced in \cite{jung2025kinetic}, and detailed in \cite{jung2026theory}.
A central difficulty of this kinetic theory is that it couples several types of variables, i.e., the physical variables and the ones directly involved in the learning process.
As a result, the derivation of closed equations for macroscopic observables requires several approximations and closures, and such technicalities may contribute to hinder the physical interpretation of the theory.
In the present work, we therefore aim at a simplified and transparent formulation of the kinetic theory for collective learning in assemblies of smart active particles, using as an example a paradigmatic active matter model, namely active Brownian particles (ABP), endowed with basic computation and communication capabilities. The main simplification in the theory results from the assumption of a full time-scale separation between the different processes at stake, and in particular between the physical dynamics and the learning process. This schematic kinetic theory allows us to present a straightforward derivation, and to address in a simple way the collective adaptation to time-dependent target states.
In addition, the presented application to the case of ABPs emphasizes the strong connection of decentralized and collective learning processes to the field of active matter.


\section{Model}

\subsection{Physical dynamics}

Agents are modeled as active Brownian particles (ABP) in two dimensions \cite{cates2013when}, located at position $\mathbf{r}$ with a self-propulsion velocity
along the direction given by an angle $\theta$.
The overdamped dynamics of an ABP with speed $v_0$ and rotational diffusion coefficient $D_R$ reads
\begin{equation}
    \mathbf{V}=\frac{d\mathbf{r}}{dt} = v_0\, \mathbf{e}(\theta), \quad \frac{d\theta}{dt} = \sqrt{2D_R}\, \xi(t),
\end{equation}
where $\mathbf{e}(\theta)$ is the unit vector in the direction $\theta$,
and $\xi(t)$ is a unit Gaussian white noise satisfying
\begin{equation}
    \langle \xi(t) \rangle = 0, \quad
    \langle \xi(t) \xi(t') \rangle = \delta(t-t').
\end{equation}
In addition to this diffusive angular dynamics, agents reorient to a given direction $\theta_0$ by stochastically jumping from $\theta$ to $\theta'=\theta_0$ at a rate $\lambda_B$. This dynamics is inspired by phototactic microswimmers like \emph{Chlamydomonas Rheinardtii} which intermittently reorient to (or against) the direction of a light source \cite{martin2016,bertin2020micro}.
Although in microswimmer experiments, the direction $\theta_0$ is related to an external light source, and is therefore the same for all microswimmers, we consider in our model that $\theta_0$ may either be the same for all agents, or take a different value for each agent.

\subsection{Learning process}

We implement learning using the same formalism as introduced in Refs.~\cite{jung2025kinetic,jung2026theory}. 

\subsubsection{Reward function}

Agents can adapt one or several of the parameters controlling their dynamics, for instance their speed $v_0$ or the angle $\theta_0$ when the latter is specific to each agent.
The set of such tunable parameters is generically called a `policy' \cite{durve2020learning,ben2023morphological,jung2025kinetic}, and is denoted as $\policy$.
Each agent adapts its policy in order to maximize a reward $\reward$.
In practice, the reward is a function $\treward(\mem)$ of a memory variable $\mem$ which is an estimate of the average value of a single-agent observable $G$. For instance, $G$ may be the instantaneous agent velocity, and the memory $\mem$ is an estimate of the average velocity, obtained by averaging the velocity over a finite time window $\tau_{\mem}$.
Each agent $i$ thus has its own observable $G_i$ and memory $\mem_i$, leading to the individual reward $\treward(\mem_i)$.
In the present work, the observable $G$ will be either the agent velocity $V_x$ along the $x$-axis, or the full velocity vector $\mathbf{V}$.

The reward function $\treward(\mem)$ typically encodes a target average state. In the case $G=V_x$, a target velocity $V_T$ along the $x$-axis is encoded in the reward function as
\begin{equation}
    \treward(\mem) = - \big( \mem - V_T\big)^2,
\end{equation}
up to an additive constant taken here to be zero without loss of generality, to lighten notations.
Generalization to a vectorial memory $\mem$ is straightforward, and will be considered later on in the paper.

\subsubsection{Memory dynamics}

The memory $\mem$ is determined by averaging the observable $G$ over a characteristic time window $\tau_{\mem}$, according to the following dynamics
\begin{equation}  \label{eq:mem:dyn}
    \frac{d\mem}{dt} = -\frac{1}{\tau_{\mem}} \big( \mem - G\big).
\end{equation}
For $t \gg \tau_{\mem}$, the solution $\mem(t)$ can approximately be written as
\begin{equation} \label{eq:memt:integral}
    \mem(t) \approx \frac{1}{\tau_{\mem}} \int_0^{\infty} dt'\, G(t-t')\, e^{-t'/\tau_{\mem}}.
\end{equation}
The memory $\mem(t)$ is thus the convolution of the observable $G(t)$ with the exponential kernel $\Theta(t)\, e^{-t/\tau_{\mem}}$, with $\Theta(t)$ the Heaviside function.

\subsubsection{Policy dynamics}

The core of the learning process lies in the policy dynamics. As mentioned above, the policy $\policy$ is a parameter (or a set of parameters) of the physical dynamics, which can be tuned to optimize the average behavior with respect to a prescribed target. 
In the present work, we consider as possible policies the individual speed $v_0$ of the agents, or the reorientation angle $\theta_0$ when the latter is independently defined for each individual agent.
The policy dynamics can be decomposed into two separate contributions, namely the teaching dynamics and random mutations.
The teaching dynamics consists in pairwise information exchange between agents, while mutations are spontaneous random policy changes, simply modeled here as a diffusion process in policy space with a diffusion coefficient $\Dmut$.

The teaching process is defined as follows. An agent interacts (i.e., exchanges information) with a neighboring agent at a given teaching rate $\lambda_T$.
At each such teaching interaction, the two selected agents $i$ and $j$ compare their rewards $\reward_i = \treward(\mem_i)$ and $\reward_j = \treward(\mem_j)$, and the agent with the highest reward has a higher probability to transfer its policy to the other agent. More precisely, agent $j$ has a probability $p_T(\reward_j,\reward_i)$ to transfer its policy $\policy_j$ to agent $i$.
The latter therefore experiences a transition from a policy $\policy_i$ to a new policy $\policy_i'=\policy_j$, for instance a change from a speed $v_{0,i}$ to $v_{0,i}'=v_{0,j}$.
The acceptance probability $p_T(\reward_j,\reward_i)$ is an increasing function of the reward difference $\reward_j-\reward_i$, taken in practice as
\begin{equation} \label{eq:def:pT}
    p_T(\reward_j,\reward_i) = \frac{1}{2} \left[ 1 +\tanh\big(\alpha_T(\reward_j-\reward_i)\big)\right],
\end{equation}
where $\alpha_T$ is a parameter.


\section{Kinetic theory}

Our goal is now to describe the statistics of the learning process for an assembly of ABP agents following the dynamics described above.
Since interactions between agents reduce to pairwise information exchange, it is natural to use a kinetic theory description \cite{de2011modeling, coscia2011mathematical, burini2016collective}.
The generic kinetic theory formalism for such a collective and decentralized learning process has been introduced in \cite{jung2025kinetic}, and further detailed in \cite{jung2026theory}.
A central difficulty of this kinetic theory is that it is not immediately formulated in terms of policy dynamics alone, since the policy dynamics depends on the reward, which is a function of the memory $\mem$, itself a time average of an observable depending on the physical variables (e.g., the agent's velocity). This coupling between different types of variables makes the theory technically more involved, and somehow conceals the relatively simple interpretation of the policy dynamics. 
Here, we consider a simplified kinetic theory for the collective learning process, obtained by assuming a full time scale separation between the physical, memory and policy dynamics. This time scale separation then allows us to formulate from the outset an approximate kinetic theory in terms of the policy degrees of freedom alone, making derivations and interpretation rather straightforward.

\subsection{Simplified description of the dynamics}

For the sake of concreteness, we now focus on the case when the policy is the particle speed, $\policy=v_0$, and the observable $G$ is chosen as the velocity component $V_x=v_0 \cos\theta$ along the $x$ direction.
The characteristic time scale of the physical dynamics can be estimated as $\tau_{\rm phys}=\max(D_R^{-1},\lambda_B^{-1})$, while the memory time scale is $\tau_{\mem}$, and the teaching time scale can be taken as $\tau_T=\lambda_T^{-1}$.
In the following, we assume the time scale separation $\tau_{\rm phys} \ll \tau_{\mem} \ll \tau_T$ to be valid.
Since $\tau_{\mem} \gg \tau_{\rm phys}$, the memory $\mem(t)$ as expressed in the integral form given in Eq.~(\ref{eq:memt:integral}) can be interpreted as a sum of a large number of statistically independent contributions, assuming that the correlation time of the physical degrees of freedom is of the same order as their relaxation time. It follows from the \emph{law of large numbers} that the probability distribution of the memory $\mem$ is sharply peaked around its steady-state average value $\langle \mem \rangle_{\rm st}$, which by virtue of Eq.~(\ref{eq:mem:dyn}) boils down to
$\langle \mem \rangle_{\rm st} = \langle V_x \rangle_{\rm st}$. Note that the steady-state average value $\langle V_x \rangle_{\rm st}$ is a function of the speed $v_0$ and no longer of the memory, so that in practice the memory $\mem$ can be replaced by a function of the policy $v_0$.
For the sake of simplicity, we also assume spatial homogeneity, so that the steady-state average observable $\langle V_x \rangle_{\rm st}$ does not depend on the position $\mathbf{r}$.
Hence under the time scale separation and spatial homogeneity assumptions, the reward function $\treward(\mem)$ can be approximated as
\begin{equation}
    \treward(\mem) \approx \treward\big( \langle V_x \rangle_{\rm st} \big) = \reward(v_0),
\end{equation}
where the last equality defines the effective reward $\reward(v_0)$, defined in terms of the policy alone. The learning process can therefore now be formulated in a closed way in terms of the policy $v_0$. We emphasize that we have chosen here the policy $\policy=v_0$ and the observable $G=V_x$ for the sake of concreteness, but the same reasoning holds for any choice of observable and policy.

\subsection{Determination of $\langle V_x \rangle_{\rm st}$}
\label{sec:stat:velocity}

To proceed further, we need an explicit determination of the steady-state average observable $\langle V_x \rangle_{\rm st}$ in terms of the policy $v_0$.
The average $\langle V_x \rangle_{\rm st}$ may be determined as follows. Due to the time-scale separation between the physical and learning dynamics, physical degrees of freedom can be assumed to evolve under a fixed policy $v_0$. In addition, physical interactions between ABPs are neglected, so that one only needs to consider the single-particle dynamics.
We thus define the single-particle probability distribution $p(\theta,\mathbf{r},t)$ of the orientation angle $\theta$ of an ABP located at position $\mathbf{r}$ at time $t$.
The distribution $p(\theta,\mathbf{r},t)$ evolves according to
\begin{equation} \label{eq:ptheta:v1}
    \frac{\partial p}{\partial t} = -\nabla \cdot \big( v_0 \mathbf{e}(\theta)\, p(\theta) \big) + D_R \frac{\partial^2 p}{\partial \theta^2} - \lambda_B p(\theta) + \lambda_B \delta(\theta-\theta_0),
\end{equation}
with $\delta$ the Dirac distribution, and where the space and time coordinates have been dropped to lighten notations.
In this first example, we consider the angle $\theta_0$ to be defined by an external field, and thus to be the same for all agents. Since the target velocity $V_T>0$ applies to the $x$-component $V_x$ of the velocity, it is natural to assume that the reorientation along $\theta_0$ align the angle $\theta$ with the $x$-axis, and we thus set $\theta_0=0$.
Under the spatial homogeneity assumption, the gradient term in Eq.~(\ref{eq:ptheta:v1}) vanishes.
As a general method, it is usually convenient to introducing the angular Fourier coefficients
\begin{equation}
    p_k = \int_{-\pi}^{\pi} d\theta \, e^{ik\theta} p(\theta)
\end{equation}
to deal with the distribution of orientations in two-dimensional active particle models \cite{peshkov2014boltzmann}.
In terms of Fourier coefficients $p_k$, Eq.~(\ref{eq:ptheta:v1}) now reads
\begin{equation}
    \frac{d p_k}{dt} = -D_R k^2 p_k + \lambda_B (1-p_k),
\end{equation}
whose stationary state is given by
\begin{equation}
    p_k^{\rm st} = \frac{\lambda_B}{k^2 D_R+\lambda_B}.
\end{equation}
To evaluate $V_x =v_0\,\mathrm{Re}(p_1)$, only the first Fourier coefficient $p_1$ is needed, and we get
\begin{equation} \label{eq:meanVx}
    \langle V_x \rangle_{\rm st} = v_0\, \mathrm{Re}(p_1^{\rm st}) = \frac{\lambda_B}{D_R+\lambda_B}.
\end{equation}

We can now explicitly evaluate the effective reward $\reward(v_0)=\treward(\langle V_x \rangle_{\rm st})$. Starting from $\treward(\mem)=-(\mem -V_T)^2$, we get 
\begin{equation} \label{eq:reward:v_0}
    \reward(v_0) = -\frac{1}{2}\reward_2 \left( v_0 - v_0^* \right)^2
\end{equation}
with
\begin{equation}
    v_0^* = \left( 1+\frac{D_R}{\lambda_B} \right) V_T, \quad \reward_2 = 2\left( \frac{\lambda_B}{D_R+\lambda_B} \right)^2. \label{eq:v0s_const}
\end{equation}
The factor $\frac{1}{2}$ in Eq.~(\ref{eq:reward:v_0}) is conventional, and introduced to match the definition used in \cite{jung2026theory}.

\subsection{Kinetic theory for collective learning}
\label{sec:kinetic}

We now explicitly derive the kinetic theory for collective learning under the assumptions of spatial homogeneity and time scale separation $\tau_{\rm phys} \ll \tau_{\mem} \ll \tau_T$. For concreteness, we write the kinetic theory in the specific case when the policy is the agent speed $v_0$, but generalization to other policies is straightforward. The kinetic theory describes the evolution of the single-agent policy density $f(v_0,t)$, normalized as $\int_0^{\infty}dv_0\, f(v_0,t)=\rho$, where $\rho=N/V$ is the real-space agent density in the homogeneous system. The policy density $f(v_0,t)$ thus describes the density of agents having a speed $v_0$. The evolution equation for $f(v_0,t)$ reads
\begin{equation} \label{eq:boltz}
    \frac{\partial f(v_0,t)}{\partial t} = \Dmut \frac{\partial^2 f}{\partial v_0^2} + I_{\rm teach}[f].
\end{equation}
The first term on the rhs of the above equation describes spontaneous policy diffusion due to random mutations with rate $\mathcal{D}$ . The term $I_{\rm teach}[f]$ describes pairwise teaching interactions between agents. Assuming that the policies of the two selected agents are statistically independent prior to the teaching interaction, the teaching term $I_{\rm teach}[f]$ can be written in an effective master equation form,
\begin{equation} \label{eq:def:Iteach}
    I_{\rm teach}[f] = \int_0^{\infty}dv_0'\, \big[ W_f(v_0|v_0') f(v_0') - W_f(v_0'|v_0) f(v_0) \big]
\end{equation}
with an effective transition rate
\begin{equation} \label{eq:def:Wf}
    W_f(v_0'|v_0) = 2\lambda_T A_c f(v_0') \, p_T\big(\reward(v_0'),\reward(v_0)\big)
\end{equation}
where the acceptance probability is given in Eq.~(\ref{eq:def:pT}).
In Eq.~(\ref{eq:def:Iteach}), $\lambda_T=\tau_T^{-1}$ is the rate of teaching interactions with a given neighbor, and $A_c$ is the area of the interaction (or communication) disk.

Several comments are in order here. First, the effective transition rate $W_f(v_0'|v_0)$ explicitly depends on $f$, at variance with standard master equations in which transition rates do not depend on the probability itself. This dependence of $W_f$ on $f$ results from the choice of treating interactions within a single-agent density description (the evolution equation for the $N$-agent probability distribution, not written here, is linear).
Second, we note that at odds with usual kinetic theory for gases, granular gases or active matter, the interaction rate is not given by a collision rate determined by the particle dynamics itself, but is rather a given parameter of the dynamics. The idea is that in a microrobot experiment, robots choose to communicate at given times with their neighbors, so that communication is not induced by a physical collision.
Last, the kinetic theory is formulated here solely in terms of the policy $v_0$, thanks to the assumptions of spatial homogeneity and time scale separation $\tau_{\rm phys} \ll \tau_{\mem} \ll \tau_T$. In contrast, the general kinetic theory for collective learning presented in \cite{jung2025kinetic,jung2026theory} was formulated jointly in terms of the physical, memory, and policy variables, making the theory more involved and less transparent.

Combining Eqs.~(\ref{eq:def:Iteach}) and (\ref{eq:def:Wf}), we get
\begin{widetext}
\begin{equation} \label{eq:Iteach:v2}
    I_{\rm teach}[f] = 2\lambda_T A_c f(v_0) \int_0^{\infty} dv_0'\, f(v_0')\big[  p_T\big(\reward(v_0),\reward(v_0')\big) - p_T\big(\reward(v_0'),\reward(v_0)\big) \big].
\end{equation}
\end{widetext}
For small $\alpha_T$, one can approximate
\begin{align} \nonumber
    p_T\big(\reward(v_0),\reward(v_0')\big) &- p_T\big(\reward(v_0'),\reward(v_0)\big)\\
    & \qquad \qquad \approx \alpha_T \big( \reward(v_0)-\reward(v_0') \big).
\end{align}
Under this approximation, we eventually obtain
\begin{equation}
    I_{\rm teach}[f] = 2\lambda_T A_c \rho f(v_0) \big( \reward(v_0) -\overline{\reward} \big)
\end{equation}
where we have defined the average reward
\begin{equation}
    \overline{\reward} = \rho^{-1} \int_0^{\infty}dv_0'\, f(v_0') \reward(v_0').
\end{equation}
Introducing the notation 
\begin{equation}
    \tilde{\lambda}_T = 2\lambda_T \alpha_T A_c,
\end{equation}
we can rewrite Eq.~(\ref{eq:boltz}) in the form
\begin{equation} \label{eq:boltz:v2}
    \frac{\partial f}{\partial t}(v_0,t) = \tilde{\lambda}_T \rho f(v_0) \big( \reward(v_0) -\overline{\reward} \big) + \Dmut \frac{\partial^2 f}{\partial v_0^2}.
\end{equation}
To proceed further, we characterize the policy distribution $\psi(v_0)=f(v_0)/\rho$ by its first two cumulants $\barv$ and $\sgv^2=\overline{v_0^2}-\barv^2$, where $\overline{v_0^n}=\int_0^{\infty}dv_0\, \psi(v_0) v_0^n$.
Note that we use here the notation $\overline{\cdots}$ for the average over the policy $v_0$, while in \cite{jung2026theory} we had to introduce a slightly different notation for policy average due to the need to distinguish different types of averages in the more involved version of the kinetic theory.

Multiplying Eq.~(\ref{eq:boltz:v2}) by $v_0$ and integrating over $v_0$, we get
\begin{equation} \label{eq:dbarv:dt}
    \frac{d\barv}{dt} = \tilde{\lambda}_T \rho \,\mathrm{Cov}\big(v_0,\reward(v_0)\big),
\end{equation}
where we have defined the covariance $\mathrm{Cov}(X,Y)=\overline{XY}-\overline{X}\,\overline{Y}$.
By a similar calculation, we also obtain
\begin{equation} \label{eq:dsqv:dt}
    \frac{d\sgv^2}{dt} = \tilde{\lambda}_T \rho \,\mathrm{Cov}\big( (v_0-\barv)^2,\reward(v_0)\big) + 2\Dmut.
\end{equation}
Eqs.~(\ref{eq:dbarv:dt}) and (\ref{eq:dsqv:dt}), which are specific limit cases of the general evolution equations derived in \cite{jung2026theory}, are valid without any restriction on the shape of the policy distribution $\psi(v_0)$ as long as it is unbounded.
Now introducing the additional assumption that $\psi(v_0)$ takes a Gaussian form, we obtain using the Stein identity (which in this case amounts to performing an integration by part)
\begin{align}
    &\mathrm{Cov}\big(v_0,\reward(v_0)\big) = \overline{\reward'(v_0)}\, \sgv^2,\\
    &\mathrm{Cov}\big( (v_0-\barv)^2,\reward(v_0)\big) = \overline{\reward''(v_0)}\, \sgv^4.
\end{align}
From the quadratic form Eq.~(\ref{eq:reward:v_0}) of $\reward(v_0)$, we get the simple expressions
\begin{equation}
    \overline{\reward'(v_0)} = -\reward_2 (\barv-v_0^*), \qquad \overline{\reward''(v_0)} = -\reward_2.
\end{equation}
We thus end up with the following closed evolution equations for $\barv$ and $\sgv^2$:
\begin{align}
    \frac{d\barv}{dt} &= -\tilde{\lambda}_T \rho \reward_2 (\barv-v_0^*)\sgv^2, \label{eq:dbarv:dt:v2}\\
    \frac{d\sgv^2}{dt} &= -\tilde{\lambda}_T \rho \reward_2 \sgv^4 + 2\Dmut \label{eq:dsqv:dt:v2}.
\end{align}
Interestingly, Eq.~(\ref{eq:dsqv:dt:v2}) admits a source term proportional to the mutation rate $\Dmut$, so that the variance $\sgv^2$ (also called diversity) converges at large time to
\begin{equation} \label{eq:sqv:infty}
    \sgv_{\infty}^2 = \left( \frac{2\Dmut}{\tilde{\lambda}_T \rho \reward_2} \right)^{1/2}.
\end{equation}
When $\sgv^2(t) \approx \sgv_{\infty}^2$, Eq.~(\ref{eq:dbarv:dt:v2}) simplifies to
\begin{equation} \label{eq:barv:dt:v3}
    \frac{d\barv}{dt} = - \frac{1}{\tau} (\barv-v_0^*),
\end{equation}
with a relaxation time $\tau$, to be interpreted as a learning time, given by
\begin{equation} \label{eq:def:tau}
    \tau = \left( 2\Dmut \tilde{\lambda}_T \rho \reward_2 \right)^{-1/2}.
\end{equation}
The expression (\ref{eq:def:tau}) of the learning time $\tau$ summarizes in a concise way the essence of the learning process.
To get a finite learning time, one needs to simultaneously include mutations ($\Dmut>0$), teaching ($\tilde{\lambda}_T>0$), a finite density of agents ($\rho>0$) and a non-flat reward ($\reward_2>0$). If any of these ingredients misses, the learning time diverges.
When $\Dmut=0$, but $\tilde{\lambda}_T \rho \reward_2>0$, a power-law relaxation $v_0-v_0^* \sim t^{-1}$ and $\sgv^2 \sim t^{-1}$ takes place \cite{jung2025kinetic,jung2026theory}. Let us emphasize that both teaching and mutations are necessary to get a finite learning time $\tau$: teaching alone, or less surprisingly mutations alone, are not enough. Note that the presence of mutations accelerates convergence, but also induces more fluctuations in the final state, since $\sgv_{\infty}^2\sim\sqrt{\Dmut}$. This is summarized in the uncertainty relation \cite{jung2025kinetic,jung2026theory},
\begin{equation} \label{eq:uncertainty}
    \tau \sgv_{\infty}^2=(\tilde{\lambda}_T \rho \reward_2)^{-1},
\end{equation}
which shows that the product $\tau \sgv_{\infty}$ is independent of the mutation rate $\Dmut$.

We have presented here a simplified treatment of the time-dependence of the solutions of Eqs.~(\ref{eq:dbarv:dt:v2}) and (\ref{eq:dsqv:dt:v2}), to focus on the physical interpretation of the result.
The form of these equations is actually generic for learning processes in arbitrary systems and policies, under the assumptions of (i) time scale separation, (ii) spatially homogeneous system and (iii) Gaussian policy distribution. The full time-dependence of the coupled Eqs.~(\ref{eq:dbarv:dt:v2}) and (\ref{eq:dsqv:dt:v2}) can be determined explicitly \cite{jung2026theory}.

\subsection{Comparison with agent-based simulations}
\label{sec:const}

We consider a constant target with the long-time goal $v_0^*=0.2.$ The policy corresponds to the active velocity $v_0$ of each agent. In particular, we investigate the learning dynamics starting from diversity $\sigma^2(t=0)=0.$ Further simulation details are given in Appendix \ref{app:simulations}.  We observe in Fig.~\ref{fig:dynamics_constant} that the time required to learn the correct policy indeed increases significantly when the mutations rate $\Dmut$ is reduced.

\begin{figure}[t]
    \centering
    \includegraphics[width=0.99\linewidth]{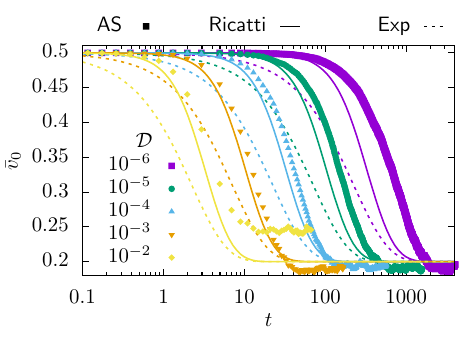}
    \caption{Learning dynamics for constant target with $v_0^*=0.2$. Data points are agent-based simulations (AS), dotted lines are a exponential decays with time scale $\tau(\Dmut)$ (Exp) and solid lines are the exact solution of Eqs.~(\ref{eq:dbarv:dt:v2}) and (\ref{eq:dsqv:dt:v2}), given in Eq.~(48) of \cite{jung2026theory}. }
    \label{fig:dynamics_constant}
\end{figure}

While 
the expression (\ref{eq:def:tau}) of $\tau$ describes qualitatively the time scale on which the learning curves decay (see dashed line in Fig.~\ref{fig:dynamics_constant}), there is a larger quantitative difference. This difference is mainly connected to the fact that assuming $\sigma^2(t) \rightarrow \sigma^2_\infty$ is constant is a quite coarse approximation, as shown by the full line in Fig.~\ref{fig:dynamics_constant} which represents the solution of the coupled Eqs.~(\ref{eq:dbarv:dt:v2}) and (\ref{eq:dsqv:dt:v2}), without approximating $\sigma^2$ as a constant.
For very small mutation rate $\Dmut$ there are additional deviations, mainly occurring due to spatial inhomogeneities which become more pronounced if learning is very slow. For very large $\Dmut$ we find that the final value for $\bar{v}_0$ actually increases. This emerges because such large $\Dmut$ do not fulfill the time scale separation assumption anymore.

\begin{figure}[t]
    \centering
    \includegraphics[width=0.99\linewidth]{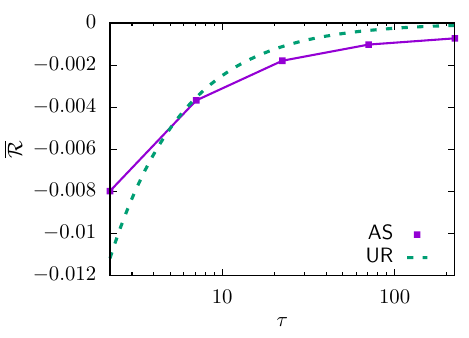}
    \caption{Average reward for constant target with $v_0^*=0.2.$ Dashed line shows the uncertainty relation (UR) derived in Eq.~(\ref{eq:UR_constant}).}
    \label{fig:time_reward_constant}
\end{figure}

We can calculate the expected average reward $\overline{\mathcal{R}}$ in the steady state. Fig.~\ref{fig:time_reward_constant} shows that it monotonically increases with the learning time $\tau.$ This connects to the uncertainty relation discussed in \cite{jung2025kinetic}, and recalled in Eq.~(\ref{eq:uncertainty}), stating that fast learning goes hand-in-hand with large fluctuations around the ideal solution, thus reducing the average reward.

We can calculate analytically the average reward $\overline{\mathcal{R}}=- \reward_2 \overline{(v_0 - v_0^*)^2}$ using the policy distribution $p(v_0) = (\sqrt{2 \pi} \sigma_\infty)^{-1} \exp(- (v_0 - v_0^*)^2/2 \sigma_\infty^2)$, yielding
\begin{equation}
    \overline{\mathcal{R}} = - \reward_2 \sigma_\infty^2
    = - \frac{\reward_2}{2 \tau \tilde \lambda_T \rho \reward_2}. \label{eq:UR_constant}
\end{equation}
There is a quite good agreement between the theoretical uncertainty relation (UR) and the agent-based simulations (AS). For small $\tau$ deviations can be explained by the missing time scale separation due to the very large large $\Dmut$. For larger $\tau$ the imperfect time scale separation between physical dynamics, memory and learning becomes relevant, leading to a tiny shift of $\bar{v}_0(t\rightarrow \infty)$. Therefore the AS results never go to $\overline{\reward} \rightarrow 0.$


\section{Time-dependent targets}
\label{sec:time_dependent}

We now turn to the case when the target velocity $V_T$ depends on time, to determine how much the learning process is able to catch up this time dependence of the target. We focus on time oscillations at angular frequency $\omega$ around a given mean value. We first consider as previously a target velocity $V_T(t)$ along the $x$-direction, before moving on to a vectorial target velocity $\mathbf{V}_T(t)$ which rotates at a constant angular velocity.

\subsection{Oscillating target velocity along the $x$-axis}
\label{sec:time_dependent_osc}

\subsubsection{Kinetic theory predictions}

We start by considering a target velocity $V_T(t)$ along the $x$-direction whose magnitude oscillates at a given angular frequency $\omega$ around a mean value $V_{T,0}$,
\begin{equation}
    V_T(t) = V_{T,0} + \Delta V_T \sin (\omega t).
\end{equation}
Consistently with the time scale separation assumption, we assume that $\omega$ corresponds to a time variation on the teaching time scale.
More precisely, we assume the time scale separation $\tau_{\mem} \ll \omega^{-1}$ on top of the previously assumed time scale separation
$\tau_{\rm phys} \ll \tau_{\mem} \ll \tau_T$.
Under these assumptions, Eqs.~(\ref{eq:dbarv:dt:v2}) and (\ref{eq:dsqv:dt:v2}) are still valid, with now a time-dependent target speed $v_0^*(t)$ given by
\begin{equation} \label{eq:timedep:v0star}
    v_0^*(t) = v_1 + v_2\sin (\omega t),
\end{equation}
with
\begin{equation}
    v_1 = \left( 1+\frac{D_R}{\lambda_B} \right) V_{T,0}, \quad
    v_2 = \left( 1+\frac{D_R}{\lambda_B} \right) \Delta V_T.
\end{equation}
Since the dynamics of $\sgv^2$ does not depend on the target policy $v_0^*(t)$, $\sgv^2(t)$ still converges to $\sgv_{\infty}^2$, as given in Eq.~(\ref{eq:sqv:infty}). At large enough time, we thus simply need to solve Eq.~(\ref{eq:barv:dt:v3}) with $v_0^*$ given by Eq.~(\ref{eq:timedep:v0star}). The solution of this equation reads
\begin{equation} \label{eq:barv:oscill}
    \barv(t) = v_1 + \frac{v_2}{\sqrt{1+(\omega \tau)^2}} \sin(\omega t-\phi)
\end{equation}
where the phase shift $\phi$ is given by
\begin{equation} \label{eq:phase:phi:v1}
    \tan \phi = \omega \tau.
\end{equation}
We thus see that the learning time $\tau$ initially introduced in the case of a time-independent target velocity $V_T$ also plays a key role in determining the phase shift between the time-dependent target velocity and oscillating collective state once a permanent regime is reached. Due to the time dependence of the target, the collective state needs to constantly adapt to the target, and therefore never reaches a state close to the target, but always follows it at some distance. The role of mutations is also essential here: in the absence of mutations (i.e., for $\Dmut=0$), one has $\tau\to \infty$, so that $\barv(t) = v_1$. In this case, the collective state does not manage to follow the time dependence of the target velocity,
and the same collective state is reached as for a time-independent target $V_T$.
In contrast, for a high mutation rate $\Dmut$ such that $\tau$ is small and $\omega \tau \ll 1$, one has
\begin{equation}
    \barv(t) \approx v_1 + v_2 \sin(\omega t-\phi) = v_0^*(t)
\end{equation}
and the collective state, as described by $\barv(t)$, follows well the time-dependent target, but at the price of a high diversity $\sigma^2$, see Eq.~(\ref{eq:sqv:infty}).

\subsubsection{Agent-based simulations}

\begin{figure}[t]
    \centering
    \includegraphics[width=0.99\linewidth]{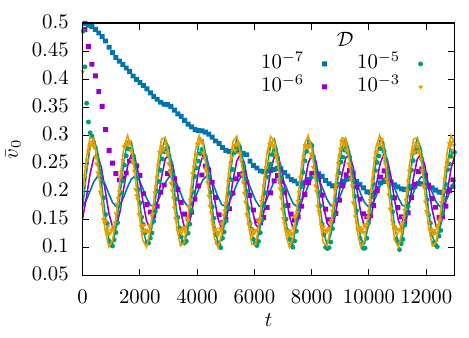}
    \caption{Learning dynamics for oscillating target with $v_0^*(t)=0.2 + 0.1 \sin(\omega t)$, $\omega=0.005$, initialized at $\bar{v}_0(t=0) = 0.5.$ Data points are agent-based simulations, full lines correspond to Eq.~(\ref{eq:barv:oscill}).}
    \label{fig:dynamics_oscillating}
\end{figure}

We simulate the agent-based model subjected to an oscillating target with $v_1=0.2$ and $v_2=0.1.$ 
We observe in Fig.~\ref{fig:dynamics_oscillating} that for very small $\tau$ adaptation is too slow to properly follow the time-dependent target, therefore, the profile becomes very flat. The solid lines are the analytical results derived in Eq.~(\ref{eq:barv:oscill}). In particular for intermediate $\tau$ (e.g., green points) the agreement is very good, predicting a slight reduction of the amplitude and a small phase shift $\phi.$

\begin{figure}[t]
    \centering
    \includegraphics[width=0.99\linewidth]{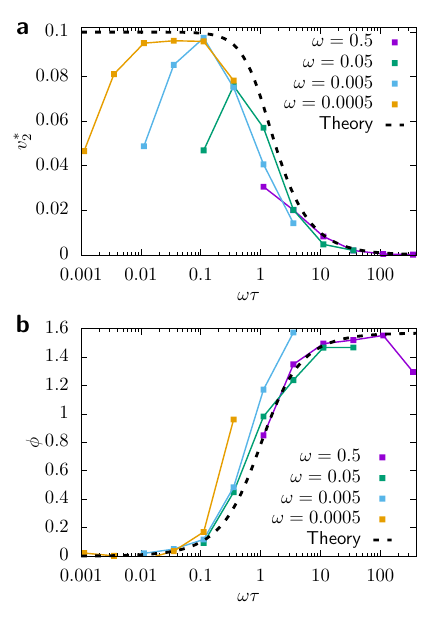}
    \caption{ Amplitude (a) and phase (b) of the average policy in time-dependent oscillating targets for various frequencies $\omega$ and mutation rates $\Dmut\in[10^{-7},10^{-2}]$. Data points are extracted from agent-based simulations, full lines correspond to Eq.~(\ref{eq:barv:oscill}) (top) and Eq.~(\ref{eq:phase:phi:v1}) (bottom), respectively.}
    \label{fig:dynamics_oscillating_ampphase}
\end{figure}

We have performed simulations as shown in Fig.~\ref{fig:dynamics_oscillating} for various frequencies of the oscillating environment $\omega$ and extracted the amplitude of the emergent oscillating mean policy and the phase compared to the external environment. We observe in Fig.~\ref{fig:dynamics_oscillating_ampphase} that both the amplitude $v_2^*$ and the phase $\phi$ collapse rather well on the theoretical master curve which only depends on $\omega \tau.$ For the amplitude we observe deviations for very large mutation rates $\Dmut,$ i.e. short learning times $\tau.$ The reason is identical to the problems observed for $\Dmut=0.01$ in Fig.~\ref{fig:dynamics_constant}.

\begin{figure}[t]
    \centering
    \includegraphics[width=0.99\linewidth]{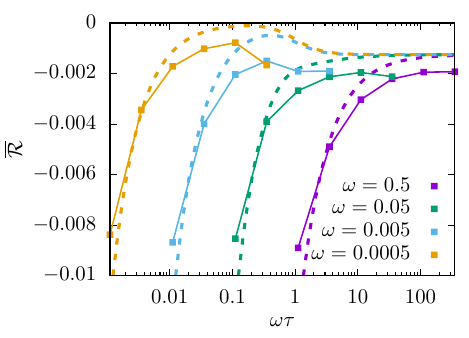}
    \caption{Average reward for oscillating target for various oscillation frequencies $\omega$. Each curves features mutation rates $\Dmut\in[10^{-7},10^{-2}].$  }
    \label{fig:time_reward_oscillating}
\end{figure}

Similar to the constant target, we calculate the average reward $\overline{\mathcal{R}}$ for the agent-based simulations, see data points in Fig.~\ref{fig:time_reward_oscillating}. For small $\omega \tau$ it follows precisely the behavior observed for the constant target. This is because learning is fast and the system can easily follow the oscillating target. However, when going to large $\tau$, it becomes increasingly difficult for the system to adapt to the time-dependent environment. As a consequence, the average reward decays for large $\tau$ and $\omega < 0.01$, featuring a maximal reward $\overline{\mathcal{R}}$. Interestingly, this maximum is not universal and we cannot extract a Master curve which only depends on $\omega \tau$, as derived for the amplitude and phase featured above. In particular, for very fast fluctuating environments it is not beneficial to have large mutation rates and adapt to the environment, but rather learn an average policy. In consequence, the maximum vanishes and the behavior is very similar to the constant target shown in Fig.~\ref{fig:time_reward_constant}.

Also in the time-dependent case we can derive an analytical expression for the average reward using the time-dependent policy distribution, $p(v_0,t) = (\sqrt{2 \pi} \sigma_\infty)^{-1} \exp(- (v_0 - \bar{v}_0(t))^2/2 \sigma_\infty^2)$ with 
$\bar{v}_0(t)$ given in Eq.~(\ref{eq:barv:oscill}). The average reward is,
\begin{align}
    \overline{\mathcal{R}} &= - \frac{ \omega \reward_2 }{2 \pi} \int_0^{2 \pi/\omega} dt \int dv_0 p(v_0,t) (v_0 - v_0^*(t))^2 \\
    &= - \reward_2 \Big( \sigma_\infty(\tau)^2 + \frac{v_2^2}{2} + \frac{v_2^2}{2(1 + (\omega \tau)^2)} - \frac{v_2^2 \cos(\phi)}{\sqrt{1 + (\omega \tau)^2}}   \Big)
\end{align}

The analytical result matches well the AS as shown in Fig.~\ref{fig:time_reward_oscillating}. In particular, both curves feature a pronounced non-monotonic behavior for $\omega < 0.01$ and a monotonic behavior for $\omega > 0.01.$

\subsection{Rotating target velocity}
\label{sec:time_dependent_rot}

We now turn to the case of a vectorial time-dependent target velocity $\mathbf{V}_T(t)$ which rotates at a fixed angular frequency $\omega$, while the 
norm $||\mathbf{V}_T||=V_T$ is kept fixed. Without loss of generality, we take the target angle $\theta_T=\omega t$, so that $\mathbf{V}_T(t)=V_T \mathbf{e}(\omega t)$.
To be able to adapt to such a rotating target velocity $\mathbf{V}_T(t)$, agents need to reorient in different directions $\theta_0$, and not only in a fixed direction determined, e.g., by an external field. We thus now consider the reorientation angle $\theta_0$ to be an internal parameter that agent can freely determine, and take as their policy. For the sake of simplicity, we now keep the speed $v_0$ identical for all agents.
The kinetic theory is expressed in terms of the single-agent policy density $f(\theta_0,t)$, and its form is similar to the one derived above in the case of the policy $v_0$, up to the $2\pi$-periodicity of the angle $\theta_0$.
However, now that the target is vectorial, a first step is to redefine the reward $\tilde{\reward}(\mem)$ as a function of a vectorial memory $\bm{\mem}$

Throughout this section, we assume the time scale separation $\tau_{\rm phys} \ll \tau_{\mem} \ll \min (\omega^{-1},\tau_T)$.

\subsubsection{Memory and reward}

We first introduce the vectorial memory $\bm{\mem}$ as the time average of the velocity $\mathbf{V}=v_0\mathbf{e}(\theta)$ over the memory time $\tau_{\mem}$, according to
\begin{equation}  \label{eq:mem:dyn:vect}
    \frac{d\bm{\mem}}{dt} = -\frac{1}{\tau_{\mem}} \big( \bm{\mem} - \mathbf{V} \big).
\end{equation}
We also need to define the reward $\tilde{\reward}(\bm{\mem})$. 
We introduce the unit vector $\mathbf{e}_T(t)=\mathbf{e}(\theta_T)$, 
where we recall that $\theta_T=\omega t$ is the orientation of the target velocity vector. We then define the reward as
\begin{equation} \label{eq:reward:vect}
    \tilde{\reward}(\bm{\mem}) = -(\bm{\mem} - V_T \mathbf{e}_T)^2.
\end{equation}
Under the time scale separation assumption $\tau_{\rm phys} \ll \tau_{\mem} \ll \min (\omega^{-1},\tau_T)$, the memory of a given agent can be expressed as the steady state average
\begin{equation} \label{eq:mem:rotating}
    \bm{\mem} = \langle v_0 \mathbf{e}(\theta) \rangle = V_0 \mathbf{e}(\theta_0),
\end{equation}
where we recall that $\theta_0$ is the agent's individual reorientation angle, and $V_0=v_0 \lambda_B/(D_R+\lambda_B)$ is the average velocity along the reorientation direction, as given in Eq.~(\ref{eq:meanVx}).
Using Eqs.~(\ref{eq:reward:vect}) and (\ref{eq:mem:rotating}), we end up with a time-dependent effective reward $\reward(\theta_0,t)$ given by
\begin{equation}
    \reward(\theta_0,t) = \reward_0 + \reward_1 \cos(\theta_0-\omega t),
\end{equation}
with $\reward_0=-(V_0^2+V_T^2)$ and $\reward_1=2 V_0 V_T$.

The evolution equation for the single-agent policy density $f(\theta_0,t)$ then reads
\begin{equation}
    \frac{\partial f(\theta_0,t)}{\partial t} = \tilde{\lambda}_T \rho \big( \reward(\theta_0,t) - \overline{\reward} \big) f(\theta_0,t)
    +\Dmut \frac{\partial^2 f}{\partial \theta_0^2}.
\end{equation}
Given that the density $\rho$ is a constant, the same equation holds for the policy distribution $\psi(\theta_0)=f(\theta_0)/\rho$,
\begin{equation} \label{eq:psi:theta0}
    \frac{\partial \psi(\theta_0,t)}{\partial t} = \tilde{\lambda}_T \rho \big( \reward(\theta_0,t) - \overline{\reward}(t) \big) \psi(\theta_0,t)
    +\Dmut \frac{\partial^2 \psi}{\partial \theta_0^2},
\end{equation}
with the average reward
\begin{equation}
    \overline{\reward}(t) = \reward_0 + \reward_1 \int_{-\pi}^{\pi}d\theta_0 \, \cos(\theta_0-\omega t) \, \psi(\theta_0).
\end{equation}

\subsubsection{Small diversity limit}
\label{sec:smalldiv:rotating}

In the small diversity limit, $\sgv \ll 2\pi$, the policy distribution $\psi(\theta_0)$ may be approximated as a sharp Gaussian distribution.
Starting from Eq.~(\ref{eq:psi:theta0}) and following similar lines are in Sec.~\ref{sec:kinetic}, we get to lowest order in $\sgth^2$,
\begin{align}
    \frac{d\barth}{dt} &= -\tilde{\lambda}_T \rho \reward_1 \sgth^2 \sin(\barth - \omega t), \label{eq:barth}\\
    \frac{d\sgth^2}{dt} &= -\tilde{\lambda}_T \rho \reward_1 \sgth^4 \cos(\barth - \omega t) + 2\Dmut \label{eq:sqth}.
\end{align}
Looking for an oscillating state $\barth=\omega t-\phi$, the stationary values of $\phi$ and $\sgth^2$ satisfy
\begin{align}
    \nu \sgth^2 \sin\phi &= \omega, \label{eq:barth:st} \\
    \nu \sgth^4 \cos\phi &= 2\Dmut, \label{eq:sqth:st}
\end{align}
with the shorthand notation
\begin{equation}
    \nu = \tilde{\lambda}_T \rho \reward_1.
\end{equation}
Combining Eqs.~(\ref{eq:barth:st}) and (\ref{eq:sqth:st}), we get closed equations for $\sgth^2$ and $\phi$:
\begin{align}
&\nu^2 \sgth^8 - \omega^2 \sgth^4 - 4\Dmut^2 = 0, \label{eq:sgth:st:closed}\\
&\sin\phi \, (\cos\phi)^{-1/2} = \omega \tau', \label{eq:phi:st:closed}
\end{align}
where we have introduced the learning time
\begin{equation} \label{eq:def:tauprime}
    \tau' = \left( 2\Dmut \tilde{\lambda}_T \rho \reward_1 \right)^{-1/2}.
\end{equation}
The definition of $\tau'$ mirrors that of $\tau$ in Eq.~(\ref{eq:def:tau}), up to the replacement of $\reward_2$ by $\reward_1$.
Solving Eq.~(\ref{eq:sgth:st:closed}), the diversity $\sgth^2$ is given by
\begin{equation}
    \sgth_{\infty}^2 = \frac{1}{\sqrt{2}} \left[ \left( \frac{\omega}{\nu} \right)^2 + \sqrt{ \left( \frac{\omega}{\nu} \right)^4 + \left( \frac{4\Dmut}{\nu}\right)^2 } \right]^{1/2}.
\end{equation}
The condition $\sgth_{\infty}^2 \ll 1$, assumed to derive Eqs.~(\ref{eq:barth}) and (\ref{eq:sqth}), therefore implies that both conditions $\omega \ll \nu$ and $\Dmut \ll \nu$ need to be satisfied.
The solution of Eq.~(\ref{eq:phi:st:closed}) in turn reads
\begin{equation} \label{eq:sin:phi:rotate}
    \sin\phi=\left( \frac{2}{1+\sqrt{1+4/(\omega\tau')^4}} \right)^{1/2}.
\end{equation}
This expression leads in particular to the following asymptotic behaviors: For $\omega\tau' \ll 1$, one has $\phi \approx \omega\tau'$, while for $\omega\tau' \gg 1$, $\phi \approx \frac{\pi}{2}-(\omega\tau')^{-2}$.

\begin{figure}[t]
    \centering
    \includegraphics[width=0.99\linewidth]{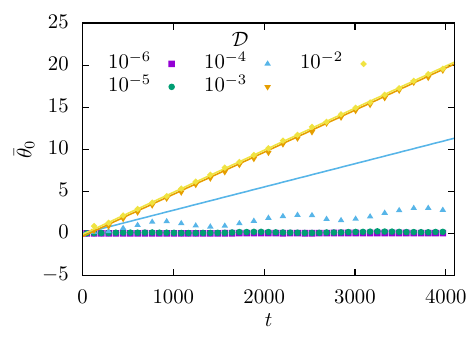}
    \caption{Learning dynamics for rotating target. Data points are results from agent-based simulations. The solid yellow and orange lines correspond to $\barth(t) = \omega t - \phi(\Dmut)$, where $\phi(\Dmut)$ is calculated from Eq.~(\ref{eq:sin:phi:rotate}). The blue solid line corresponds to Eq.~(\ref{eq:omega_unlock}). }
    \label{fig:dynamics_rotating}
\end{figure}

We study numerically in the agent-based model a rotating target with $\omega=0.005.$ 
The learning dynamics are shown in Fig.~\ref{fig:dynamics_rotating}. 
For large enough mutation rate $\Dmut$, the angle follows the rotating target, therefore the average $\barth$ increases linearly with slope $\omega$.
This result is consistent with the analytical prediction $\barth=\omega t-\phi$, with $\phi$ given by Eq.~(\ref{eq:sin:phi:rotate}),
as shown by the full lines in Fig.~\ref{fig:dynamics_rotating}.
By contrast, for very small mutation rate $\Dmut$, the target rotation is too fast compared to the learning time $\tau'$ and the average angle $\barth$ stays constant. In the intermediate regime, we observe that $\barth$ tries to follow the rotating target and increases linearly on average, with oscillations around the average slope. Yet this linear increase is too slow to catch up the target (Fig.~\ref{fig:dynamics_rotating}).

\subsubsection{Unlocked phase}

\begin{figure}[t]
    \centering
    \includegraphics[width=0.99\linewidth]{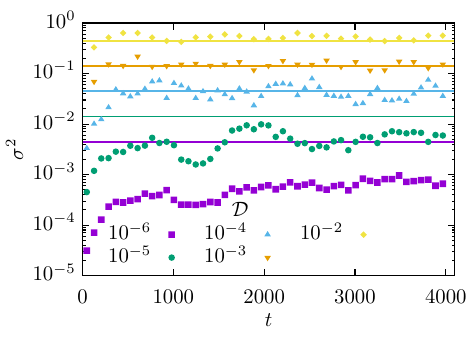}
    \caption{Variance $\sgth^2$ as a function of time in the agent-based model. Horizontal lines indicate the approximate prediction $\sgth_{\infty}^2$ for the plateau value, given in Eq.~(\ref{eq:sgth:approx:rotat}).}
    \label{fig:variance}
\end{figure}

Although the analytical solution $\barth=\omega t-\phi$, with $\phi$ given by Eq.~(\ref{eq:sin:phi:rotate}), is observed in the numerical simulations for relatively large enough values of the mutation rate $\Dmut$, it does not correctly describe the numerical observations for intermediate values of $\Dmut$, where the collective behavior no longer follows the target. Such an unlocked state may be qualitatively understood as follows.
Let us now consider $\phi=\omega t-\barth$ as a dynamical variable. In terms of $\phi$, Eqs.~(\ref{eq:barth}) and (\ref{eq:sqth}) read
\begin{align}
    &\frac{d\phi}{dt} = \omega - \nu \sgth^2 \sin\phi, \label{eq:barth:phi} \\
    &\frac{d\sgth^2}{dt} = -\nu \sgth^4 \cos\phi + 2\Dmut. \label{eq:sqth:phi}
\end{align}
Numerical simulations of the agent based model show that $\sgth^2$ oscillates around a plateau value, see Fig.~\ref{fig:variance}.
This plateau value $\sgth_{\infty}^2$ may be approximately predicted by neglecting the $\cos\phi$ term in Eq.~(\ref{eq:sqth:phi}), yielding
\begin{equation} \label{eq:sgth:approx:rotat}
    \sgth_{\infty}^2 = \sqrt{\frac{2\Dmut}{\nu}}.
\end{equation}

Although this is a priori a rough approximation, it is seen to predict rather well the plateau values, as shown in Fig.~\ref{fig:variance} where the values given by Eq.~(\ref{eq:sgth:approx:rotat}) are plotted as horizontal lines. Significant deviations emerge in the unlocked phase where $\cos\phi$ yields a significant contribution. Neverthess, we expect that the transition is described rather accuratly by this approximation.
One then gets from Eq.~(\ref{eq:barth:phi}),
\begin{equation} \label{eq:barth:phi:approx}
    \frac{d\phi}{dt} = \omega - \sqrt{2\Dmut\nu} \,\sin\phi.
\end{equation}
Equations of this kind a known from locking phenomena in oscillators, in particular the Adler equation \cite{adler1946study}, which have been analytical analyzed in great detail \cite{shlomovitz2014phase,rusch2026intermediate}. When $\sqrt{2\Dmut\nu}>\omega$, $\phi$ converges to a fixed point, as described in Sec.~\ref{sec:smalldiv:rotating} (where the $\cos\phi$ term was explicitly taken into account).
In contrast, when $\sqrt{2\Dmut\nu}<\omega$, Eq.~(\ref{eq:barth:phi:approx}) has no fixed point solution, and $\phi$ continuously increases with time.
The time derivative $\dot\phi=d\phi/dt>0$ has an average value $\langle \dot\phi \rangle =2\pi/T$, where the time period $T=\int d\phi/\dot\phi$ is given by
\begin{equation}
    T = \int_{-\pi}^{\pi} \frac{d\phi}{\sqrt{\omega - \sqrt{2\Dmut\nu} \,\sin\phi}} = \frac{2\pi}{\sqrt{\omega^2-2\Dmut\nu}},
\end{equation}
leading to
\begin{align}\label{eq:omega_unlock}
    \left\langle \frac{d\barth}{dt} \right\rangle = \omega - \sqrt{\omega^2-2\Dmut\nu}  \qquad (\sqrt{2\Dmut\nu}<\omega).
\end{align}
This solution corresponds to the unlocked collective state observed on Fig.~\ref{fig:dynamics_rotating}, where the system tries to follow the target, but does not manage to adapt fast enough. While visually there is a large deviation between the analytical prediction in Eq.~(\ref{eq:omega_unlock}) for $\mathcal{D}=10^{-4}$ and the agent-based simulations, this deviation emerges since the transition is very sensitive to the mutation rate $\mathcal{D}.$ 

\begin{figure}[t]
    \centering
    \includegraphics[width=0.99\linewidth]{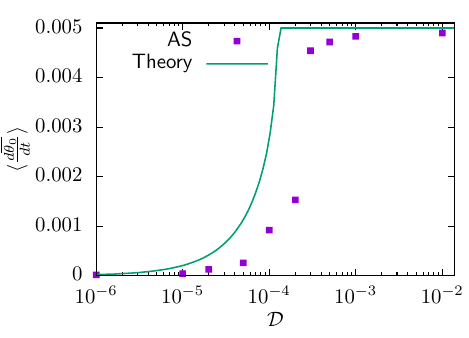}
    \caption{Unlocking transition in the collective learning model for a rotating target with $\omega=0.005$. The figure shows the time-averaged change of the average policy $\barth$ for various mutation rates $\mathcal{D}$. Data points are agent-based simulations (AS) and full line is the analytical prediction Eq.~(\ref{eq:barth:phi:approx}).  }
    \label{fig:transition}
\end{figure}

We therefore explicitly investigate the unlocking transition in Fig.~\ref{fig:transition}. We observe that the agent-based model features an unlocking transition as predicted by the analytical results in Eq.~(\ref{eq:barth:phi:approx}). In particular, when approaching the critical mutation rate  $\mathcal{D}_c = \omega^2 / 2 \nu$ from above, the  time-averaged change of the average policy $\barth$ becomes very sensitive to $\mathcal{D}.$ The only difference between theory and simulations is a shift in the critical mutation rate  $\mathcal{D}_c$ which emerges due to the approximation in the kinetic theory, in particular the assumption of a perfect time-scale separation which is not given in the agent-based simulations. This leads to increased fluctuations in the agent-based model which make it more difficult to synchronize with the target. In consequence, larger mutations rates are required and $\mathcal{D}_c$ shifts to the right.

\subsubsection{Large diversity limit}

Finally, we briefly consider the opposite limit of a large diversity $\sgth^2$, in which case the policy distribution $\psi(\theta_0)$ becomes almost uniform over the interval $(-\pi,\pi]$. The policy distribution is then conveniently described in terms of low order angular Fourier modes,
\begin{equation}
    \psi_k = \int_{-\pi}^{\pi} d\theta_0 \, e^{ik\theta_0} \psi(\theta_0).
\end{equation}
One has in particular $\psi_0=1$ due to the normalization condition.
Note that the present angular Fourier mode expansion of the policy distribution $\psi(\theta_0)$ should not be confused with the Fourier mode expansion of the distribution $p(\theta)$ of the physical angle $\theta$ of the particle, considered in Sec.~\ref{sec:stat:velocity}.
Expanding Eq.~(\ref{eq:psi:theta0}) in angular Fourier series, we obtain
\begin{align}
    \frac{d\psi_k}{dt} &= \frac{\nu}{2} e^{-i\omega t} \big( \psi_{k+1}-\psi_1\psi_k \big) \nonumber \\
    & \qquad \qquad + \frac{\nu}{2} e^{i\omega t} \big( \psi_{k-1}-\psi_{-1}\psi_k \big) -\Dmut k^2 \psi_k.
\end{align}
We assume that the distribution $\psi(\theta_0)$ follows the rotating target velocity as $\psi(\theta_0)=\tilde{\psi}(\theta_0-\omega t)$, so that $\psi_k=e^{ik\omega t}\tilde{\psi}_k$.
Once $\tilde{\psi}$ has reached a stationary state, the Fourier coefficients $\tilde{\psi}_k$ satisfy
\begin{equation} \label{eq:tildepsi:k}
    ik\omega \tilde{\psi}_k = \nu \big( \tilde{\psi}_{k+1}-\tilde{\psi}_1 \tilde{\psi}_k + \tilde{\psi}_{k-1}-\tilde{\psi}_{-1}\tilde{\psi}_k \big) -\Dmut k^2 \tilde{\psi}_k.
\end{equation}
In the following, we assume that $\tilde{\psi}_1 \ll 1$, and that $\tilde{\psi}_k \ll \tilde{\psi}_1$. At linear order in $\tilde{\psi}_1$, Eq.~(\ref{eq:tildepsi:k}) boils down to
\begin{equation}
    i\omega \tilde{\psi}_1 = \nu - \Dmut \tilde{\psi}_1,
\end{equation}
whose solution reads
\begin{equation}
    \tilde{\psi}_1 = \frac{\nu}{2(\Dmut + i\omega)}.
\end{equation}
The assumption that the modulus
\begin{equation}
    |\tilde{\psi}_1| = \frac{\nu}{2\sqrt{\Dmut^2 + \omega^2}}
\end{equation}
is small is thus valid either when $\Dmut \gg \nu$ or $\omega \gg \nu$, which is precisely the opposite regime as the one found for the sharp distribution case, which therefore outlines the consistency of our approach.
The phase $\phi$ is given by
\begin{equation}
    \tan \phi = \frac{\omega}{\Dmut}.
\end{equation}
We thus observe that for both small and large diversity limits, the phase shift $\phi$ is an increasing function of the angular frequency $\omega$ of the rotating target velocity, ranging from
$\phi\approx 0$ at low $\omega$ to $\phi \approx \frac{\pi}{2}$ at large $\omega$. However, the characteristic frequency scale $\omega_0$ around which the crossover from $\phi\approx 0$ to $\phi \approx \frac{\pi}{2}$ occurs may strongly differ between the small and large diversity cases. For a small diversity, $\omega_0 \sim ( \Dmut \tilde{\lambda}_T \rho \reward_1)^{1/2}$,
and both mutations and teaching concur to determine the phase lag $\phi$. In constrast, for a large diversity, $\omega_0 \sim \Dmut$, and the phase lag is purely determined by mutations.
In this case, teaching only determines the amplitude $|\tilde{\psi}_1|$, so that in the absence of teaching, $|\tilde{\psi}_1|=0$ as expected.

\section{Discussion}

In this work, we have shown how simple active agents able to tune their dynamics can collectively adapt to a prescribed target through a decentralized learning process.
The average target state is encoded into an individual reward function, which under a time scale separation assumption can be expressed in terms of the policy (i.e., the tunable parameter of the dynamics). By exchanging information on their respective policies and rewards, agent can collectively adapt their dynamics to optimize their reward.
A diffusion noise, called mutations by analogy with evolutionary biological processes, additionally generates a spontaneous and stochastic evolution of policy.

A kinetic theory allows us to describe the evolution of the policy distribution under the adaptation process, for both time-independent and time-dependent targets. 
In the case of time-independent target states, the policy distribution across the population of agents converges to a peaked distribution around the target state.
The amplitude of fluctuations around the target state are determined by the mutation rate. A higher mutation rate induces a faster adaptation process, but also larger fluctuations around the target in the final state. The adaptation process can be characterized by the learning time $\tau$, given in Eq.~(\ref{eq:def:tau}), which combines the teaching rate and the mutation rate, showing that both mechanisms play an equally important role in the adaptation process.

For time-oscillating target states at angular frequency $\omega$, adaptation is a perpetual process, and the collective state follows the target with a phase shift determined by $\omega \tau'$, where $\tau'$ is defined in Eq.~(\ref{eq:def:tauprime}) and has the same interpretation as $\tau$. A higher mutation rate leads to a lower phase shift, but also to a weaker amplitude of the collective response to the oscillating target. In the case of the rotating vectorial target velocity, an unlocking transition is observed for intermediate values of the mutation rate, and the collective state rotates but does not manage to adapt fast enough to follow the target.

As for future work, it would be of interest to expand the present ABP model to include physical interactions, like steric or alignment interactions, which are expected to play an important role in dense groups of microrobots, in the spirit of morphological computations \cite{pfeifer2009morphological,ben2023morphological}. Another important issue would be to go beyond the strict time-scale separation assumed here, notably to be able to couple the adaptation (or learning) dynamics to spatial heterogeneities. The general kinetic theory formalism in the case of non-separated time scales has been presented in \cite{jung2026theory}, and applied to some elementary models.
The application of this general formalism to standard active matter models like ABP or Run-and-Tumble Particles (RTP) would be a natural next step.

\subsection*{Acknowledgments} The authors are grateful to O.~Dauchot and M. Ozawa for useful discussions. This research was partly funded by the Austrian Science Fund (FWF) 10.55776/PAT1139125.

\appendix

\section{Agent-based simulations}
\label{app:simulations}

We have performed agent-based simulations throughout this work to confirm and illustrate analytical results obtained from kinetic theory. Generally, these simulations follow the same algorithm as described in Ref.~\cite{jung2026theory}. 

We use $N=1000$ agents in a square box of size $L=10$, leading to an agent density of $\rho=10.$ The rotational diffusion coefficient of the active agents is set to $D_R=50$. Similarly, the jumping rate to the direction $\theta_0$ of the light source is also set to $\lambda_B=50.$ In the analytical results we have assumed a perfect time scale separation between physical dynamics, memory and learning. In agent-based simulations we can of course not fulfill this condition, we try to separate the scales as much as possible. We choose $\lambda_\mem = 5$, $\lambda_T=0.1$ and $\alpha_T=10. $ While this leads to results which compare very well to the analytics, there are a few differences, in particular when going to very high mutation rates, as discussed in the main text.

For the constant and time-dependent oscillating target the policy which agents try to optimize is the active velocity $v_0. $ The constant target in Sec.~\ref{sec:const} is given by $V_T = 0.1$ and $\theta_0=0.$ Following Eq.~(\ref{eq:v0s_const}) we therefore have $v_0^* = 0.2,$ as stated in the main text. The oscillating target in Sec.~\ref{sec:time_dependent_osc} is characterized by $V_{T,0} = 0.1$ and $\Delta V_T=0.05$, which leads to the $v_0^*(t)$ as given in Eq.~(\ref{eq:timedep:v0star}).

To adapt to the rotating target the active velocity $v_0=0.1$ is kept constant and instead, the policy is now given by the direction $\theta_0$ of the light source. The rotation frequency is given by $\omega=0.005.$

\bibliographystyle{apsrev4-2}
\bibliography{biblio}

@article{ariosto2025replication,
  title={Replication and Information Extraction in a Minimal Agent-Environment Model},
  author={Ariosto, Sebastiano and Garnier-Brun, Jerome and Saglietti, Luca and Straziota, Davide},
  journal={arXiv preprint arXiv:2509.23212},
  year={2025}
}

@incollection{pfeifer2009morphological,
  title={Morphological computation--connecting brain, body, and environment},
  author={Pfeifer, Rolf and G{\'o}mez, Gabriel},
  booktitle={Creating brain-like intelligence: From basic principles to complex intelligent systems},
  pages={66--83},
  year={2009},
  publisher={Springer}
}

@article{ben2023morphological,
	title={Morphological computation and decentralized learning in a swarm of sterically interacting robots},
	author={Ben Zion, Matan Yah and Fersula, Jeremy and Bredeche, Nicolas and Dauchot, Olivier},
	journal={Science Robotics},
	volume={8},
	number={75},
	pages={eabo6140},
	year={2023},
	publisher={American Association for the Advancement of Science},
	doi = {10.1126/scirobotics.abo6140}
}

@article{bredeche2022social,
	author = {Bredeche, Nicolas and Fontbonne, Nicolas},
	title = {Social learning in swarm robotics},
	journal = {Philosophical Transactions of the Royal Society B: Biological Sciences},
	volume = {377},
	number = {1843},
	pages = {20200309},
	year = {2022},
	doi = {10.1098/rstb.2020.0309},
	URL = {https://royalsocietypublishing.org/doi/abs/10.1098/rstb.2020.0309}
}

@misc{cazenille2024signaling,
	title={Signaling and Social Learning in Swarms of Robots}, 
	author={Leo Cazenille and Maxime Toquebiau and Nicolas Lobato-Dauzier and Alessia Loi and Loona Macabre and Nathanael Aubert-Kato and Anthony Genot and Nicolas Bredeche},
	year={2024},
	eprint={2411.11616},
	archivePrefix={arXiv},
	primaryClass={cs.RO},
	url={https://arxiv.org/abs/2411.11616}, 
}

@misc{fersula2026aggregating,
      title={Aggregating swarms through morphology handling design contingencies: from the sweet spot to a rich expressivity}, 
      author={Jeremy Fersula and Nicolas Bredeche and Olivier Dauchot},
      year={2026},
      eprint={2601.07610},
      archivePrefix={arXiv},
      primaryClass={cond-mat.soft},
      url={https://arxiv.org/abs/2601.07610}, 
}

@article{ramaswamy2010mechanics,
	author = "Ramaswamy, Sriram",
	title = "The Mechanics and Statistics of Active Matter", 
	journal= "Annual Review of Condensed Matter Physics",
	year = "2010",
	volume = "1",
	number = "Volume 1, 2010",
	pages = "323-345",
	doi = "10.1146/annurev-conmatphys-070909-104101",
	url = "https://www.annualreviews.org/content/journals/10.1146/annurev-conmatphys-070909-104101",
	publisher = "Annual Reviews",
	issn = "1947-5462",
	type = "Journal Article",
}

@article{ramaswamy2019active,
	title = {Active fluids},
	author = {Ramaswamy, S.},
	journal = {Nat. Rev. Phys.},
	volume = {1},
	pages = {640–642},
	year = {2019},
	doi = {10.1038/s42254-019-0120-9}
}

@article{marchetti2013hydrodynamics,
	title = {Hydrodynamics of soft active matter},
	author = {Marchetti, M. C. and Joanny, J. F. and Ramaswamy, S. and Liverpool, T. B. and Prost, J. and Rao, Madan and Simha, R. Aditi},
	journal = {Rev. Mod. Phys.},
	volume = {85},
	issue = {3},
	pages = {1143--1189},
	numpages = {0},
	year = {2013},
	month = {Jul},
	publisher = {American Physical Society},
	doi = {10.1103/RevModPhys.85.1143},
	url = {https://link.aps.org/doi/10.1103/RevModPhys.85.1143}
}

@article{bechinger2016active,
  title = {Active particles in complex and crowded environments},
  author = {Bechinger, Clemens and Di Leonardo, Roberto and L\"owen, Hartmut and Reichhardt, Charles and Volpe, Giorgio and Volpe, Giovanni},
  journal = {Rev. Mod. Phys.},
  volume = {88},
  issue = {4},
  pages = {045006},
  numpages = {50},
  year = {2016},
  month = {Nov},
  publisher = {American Physical Society},
  doi = {10.1103/RevModPhys.88.045006},
  url = {https://link.aps.org/doi/10.1103/RevModPhys.88.045006}
}

@article{kaspar2021rise,
	title={The rise of intelligent matter},
	author={Kaspar, Corinna and Ravoo, Bart Jan and van der Wiel, Wilfred G. and Wegner, Seraphine V. and Pernice, Wolfram H. P.},
	journal={Nature},
	volume={594},
	number={7863},
	pages={345--355},
	year={2021},
	publisher={Nature Publishing Group UK London},
	doi = {10.1038/s41586-021-03453-y}
}

@article{levine2023physics,
	author = {Levine, Herbert and Goldman, Daniel I.},
	title = {Physics of smart active matter: integrating active matter and control to gain insights into living systems},
	journal = {Soft Matter},
	year  = {2023},
	volume  = {19},
	issue  = {23},
	pages  = {4204-4207},
	publisher  = {The Royal Society of Chemistry},
	doi  = {10.1039/D3SM00171G},
	url  = {http://dx.doi.org/10.1039/D3SM00171G}
}

@article{cichos2020machine,
	title = {Machine learning for active matter},
	author = {Cichos, F. and Gustavsson, K. and Mehlig, B. and Volpe, G.},
	journal = {Nat. Mach. Intell.},
	volume = {2},
	pages = {94–103},
	year = {2020},
	doi = {10.1038/s42256-020-0146-9}
}

@article{baulin2025intelligent,
    title = {Intelligent soft matter: towards embodied intelligence},
    author = {Baulin, Vladimir A. and Giacometti, Achille and Fedosov, Dmitry A. and Ebbens, Stephen and Varela-Rosales, Nydia R. and Feliu, Neus and Chowdhury, Mithun and Hu, Minghan and Füchslin, Rudolf and Dijkstra, Marjolein and Mussel, Matan and van Roij, René and Xie, Dong and Tzanov, Vassil and Zu, Mengjie and Hidalgo-Caballero, Samuel and Yuan, Ye and Cocconi, Luca and Ghim, Cheol-Min and Cottin-Bizonne, Cécile and Miguel, M. Carmen and Esplandiu, Maria Jose and Simmchen, Juliane and Parak, Wolfgang J. and Werner, Marco and Gompper, Gerhard and Hanczyc, Martin M.},
    journal = {Soft Matter},
    volume = {21},
    number = {21},
    pages = {4129-4145},
    year = {2025},
    doi = {10.1039/d5sm00174a},
    url = {https://doi.org/10.1039/d5sm00174a},
}

@article{vansaders2023informational,
  title={Measurement-induced phase transitions in informational active matter},
  author={VanSaders, Bryan and Fruchart, Michel and Vitelli, Vincenzo},
  journal={PNAS Nexus},
  volume={5},
  number={4},
  pages={pgag077},
  year={2026},
  publisher={Oxford University Press US}
}

@misc{cocconi2024dissipation,
	title={Dissipation-accuracy tradeoffs in autonomous control of smart active matter}, 
	author={Cocconi, L. and Mahault, B. and Piro, L.},
	year={2024},
	eprint={2409.12595},
	archivePrefix={arXiv},
	primaryClass={cond-mat.stat-mech},
	url={https://arxiv.org/abs/2409.12595}, 
}

@article{ziepke2022multi,
author = {Ziepke, A. and Maryshev, I. and Aranson, I.S. and Frey, E.},
title = {Multi-scale organization in communicating active matter},
journal = {Nat. Commun.},
volume = {13},
pages = {6727},
year = {2022},
doi = {https://doi.org/10.1038/s41467-022-34484-2},
}

@article{negi2025binary,
title = {Binary mixtures of intelligent active Brownian particles with visual perception},
author = {Negi, Rajendra Singh and G Winkler, Roland and Gompper, Gerhard},
journal = {New Journal of Physics},
year = {2025},
volume = {27},
number = {10},
pages = {103301},
doi = {10.1088/1367-2630/ae140c},
url = {https://doi.org/10.1088/1367-2630/ae140c},
}

@article{goh2026adaptative,
author = {Goh, Segun and Haustein, Dennis and Gompper, Gerhard},
title = {Adaptive Control of Run-and-Tumble Escape in Pursuit-Evasion Dynamics of Intelligent Active Particles},
journal = {Advanced Intelligent Systems},
volume = {8},
number = {5},
pages = {e202500852},
doi = {https://doi.org/10.1002/aisy.202500852},
url = {https://advanced.onlinelibrary.wiley.com/doi/abs/10.1002/aisy.202500852},
year = {2026}
}

@article{iyer2026emergent,
	author={Iyer, Priyanka and Goh, Segun and Gompper, Gerhard},
	title={Emergent self-organisation of intelligent active particles},
	journal={Europhysics Letters},
	url={http://iopscience.iop.org/article/10.1209/0295-5075/ae9db0},
	year={2026},
    note={in press},
}

@article{zampetaki2021collective,
author = {Alexandra V. Zampetaki  and Benno Liebchen  and Alexei V. Ivlev  and Hartmut Löwen },
title = {Collective self-optimization of communicating active particles},
journal = {Proceedings of the National Academy of Sciences},
volume = {118},
number = {49},
pages = {e2111142118},
year = {2021},
doi = {10.1073/pnas.2111142118},
URL = {https://www.pnas.org/doi/abs/10.1073/pnas.2111142118},
}

@misc{munoz2026emergent,
      title={Emergent aggregation from collective foraging}, 
      author={Gorka Mu\~noz-Gil and Andrea L\'opez-Incera and Vide Ramsten and Giovanni Volpe and Thomas M\"uller and Hans J. Briegel},
      year={2026},
      eprint={2608.28046},
      archivePrefix={arXiv},
      primaryClass={cond-mat.stat-mech},
      url={https://arxiv.org/abs/2608.28046}, 
}

@article{majidi2019soft,
	author = {Majidi, Carmel},
	title = {Soft-Matter Engineering for Soft Robotics},
	journal = {Advanced Materials Technologies},
	volume = {4},
	number = {2},
	pages = {1800477},
	doi = {10.1002/admt.201800477},
	url = {https://onlinelibrary.wiley.com/doi/abs/10.1002/admt.201800477},
	year = {2019}
}

@article{goldman2024robot,
	author = {Goldman, D. I. and Rocklin, D. Z.},
	title = {Robot swarms meet soft matter physics},
	journal = {Science Robotics},
	volume = {9},
	number = {86},
	pages = {eadn6035},
	year = {2024},
	doi = {10.1126/scirobotics.adn6035},
	URL = {https://www.science.org/doi/abs/10.1126/scirobotics.adn6035}
}

@article{pishvar2020foundations,
	author = {Pishvar, Maya and Harne, Ryan L.},
	title = {Foundations for Soft, Smart Matter by Active Mechanical Metamaterials},
	journal = {Advanced Science},
	volume = {7},
	number = {18},
	pages = {2001384},
	doi = {10.1002/advs.202001384},
	url = {https://onlinelibrary.wiley.com/doi/abs/10.1002/advs.202001384},
	year = {2020}
}

@article{liebchen2019optimal,
	doi = {10.1209/0295-5075/127/34003},
	url = {https://dx.doi.org/10.1209/0295-5075/127/34003},
	year = {2019},
	month = {sep},
	publisher = {EDP Sciences, IOP Publishing and Società Italiana di Fisica},
	volume = {127},
	number = {3},
	pages = {34003},
	author = {Liebchen, B. and L\"{o}wen, H.},
	title = {Optimal navigation strategies for active particles},
	journal = {Europhysics Letters}
}

@article{nasiri2023optimal,
	doi = {10.1209/0295-5075/acc270},
	url = {https://dx.doi.org/10.1209/0295-5075/acc270},
	year = {2023},
	month = {mar},
	publisher = {EDP Sciences, IOP Publishing and Società Italiana di Fisica},
	volume = {142},
	number = {1},
	pages = {17001},
	author = {Nasiri, M. and L\"{o}wen, H. and Liebchen, B.},
	title = {Optimal active particle navigation meets machine learning},
	journal = {Europhysics Letters}
}

@article{piro2021optimal,
	title = {Optimal navigation strategies for microswimmers on curved manifolds},
	author = {Piro, Lorenzo and Tang, Evelyn and Golestanian, Ramin},
	journal = {Phys. Rev. Res.},
	volume = {3},
	issue = {2},
	pages = {023125},
	numpages = {9},
	year = {2021},
	month = {May},
	publisher = {American Physical Society},
	doi = {10.1103/PhysRevResearch.3.023125},
	url = {https://link.aps.org/doi/10.1103/PhysRevResearch.3.023125}
}

@article{piro2022optimal,
	doi = {10.1088/1367-2630/ac9079},
	url = {https://dx.doi.org/10.1088/1367-2630/ac9079},
	year = {2022},
	month = {sep},
	publisher = {IOP Publishing},
	volume = {24},
	number = {9},
	pages = {093037},
	author = {Piro, L. and Mahault, B. and Golestanian, R.},
	title = {Optimal navigation of microswimmers in complex and noisy environments},
	journal = {New Journal of Physics}
}

@article{piro2022efficiency,
	title = {Efficiency of navigation strategies for active particles in rugged landscapes},
	author = {Piro, L. and Golestanian, R. and Mahault, B.},
	journal = {Front. Phys.},
	volume = {10},
	pages = {1034267},
	year = {2022},
	doi = {10.3389/fphy.2022.1034267}
}

@article{borra2021optimal,
	doi = {10.1088/1742-5468/ac12c6},
	url = {https://dx.doi.org/10.1088/1742-5468/ac12c6},
	year = {2021},
	month = {aug},
	publisher = {IOP Publishing and SISSA},
	volume = {2021},
	number = {8},
	pages = {083401},
	author = {Borra, F. and Cencini, M. and Celani, A.},
	title = {Optimal collision avoidance in swarms of active Brownian particles},
	journal = {Journal of Statistical Mechanics: Theory and Experiment}
}

@article{yang2022autonomous,
	title={Autonomous environment-adaptive microrobot swarm navigation enabled by deep learning-based real-time distribution planning},
	author={Yang, Lidong and Jiang, Jialin and Gao, Xiaojie and Wang, Qinglong and Dou, Qi and Zhang, Li},
	journal={Nature Machine Intelligence},
	volume={4},
	number={5},
	pages={480--493},
	year={2022},
	publisher={Nature Publishing Group UK London},
	doi = {10.1038/s42256-022-00482-8}
}

@article{bilai2026theory,
  title={Optimal navigation in stochastic and disordered gridworlds},
  author={Bila\"{\i} Biloa, K\'evin and Pierre-Louis, Olivier},
  journal={arXiv preprint arXiv:2605.03568},
  year={2026}
}

@misc{olsen2026information,
      title={Information bound on navigation speed in smart active matter}, 
      author={Kristian Stølevik Olsen and Mitsusuke Tarama and Hartmut Löwen},
      year={2026},
      eprint={2602.23988},
      archivePrefix={arXiv},
      url={https://arxiv.org/abs/2602.23988}, 
}

@article{monderkamp2022active,
author = {Monderkamp, Paul A and Schwarzendahl, Fabian Jan and Klatt, Michael A and Löwen, Hartmut},
title = {Active particles using reinforcement learning to navigate in complex motility landscapes},
doi = {10.1088/2632-2153/aca7b0},
url = {https://doi.org/10.1088/2632-2153/aca7b0},
year = {2022},
month = {dec},
publisher = {IOP Publishing},
volume = {3},
number = {4},
pages = {045024},
journal = {Machine Learning: Science and Technology},
}

@article{bonnemain2023pedestrians,
	title = {Pedestrians in static crowds are not grains, but game players},
	author = {Bonnemain, Thibault and Butano, Matteo and Bonnet, Th\'eophile and Echeverr\'{\i}a-Huarte, I\~naki and Seguin, Antoine and Nicolas, Alexandre and Appert-Rolland, C\'ecile and Ullmo, Denis},
	journal = {Phys. Rev. E},
	volume = {107},
	issue = {2},
	pages = {024612},
	numpages = {12},
	year = {2023},
	month = {Feb},
	publisher = {American Physical Society},
	doi = {10.1103/PhysRevE.107.024612},
	url = {https://link.aps.org/doi/10.1103/PhysRevE.107.024612}
}

@article{echevarria2023body,
	title = {Body and mind: Decoding the dynamics of pedestrians and the effect of smartphone distraction by coupling mechanical and decisional processes},
	journal = {Transportation Research Part C: Emerging Technologies},
	volume = {157},
	pages = {104365},
	year = {2023},
	issn = {0968-090X},
	doi = {10.1016/j.trc.2023.104365},
	url = {https://www.sciencedirect.com/science/article/pii/S0968090X23003558},
	author = {Echeverr\'{\i}a-Huarte, I. and Nicolas, A.}
}

@article{mo2023challenges,
	title = {Challenges and attempts to make intelligent microswimmers},
	author = {Mo C., Li G. and Bian X.},
	journal = {Front. Phys.},
	volume = {11},
	pages = {1279883},
	year = {2023},
	doi = {10.3389/fphy.2023.1279883}
}

@article{boccardo2024reinforcement,
  title = {Reinforcement learning with thermal fluctuations at the nanoscale},
  author = {Boccardo, Francesco and Pierre-Louis, Olivier},
  journal = {Phys. Rev. E},
  volume = {110},
  issue = {2},
  pages = {L023301},
  numpages = {6},
  year = {2024},
  month = {Aug},
  publisher = {American Physical Society},
  doi = {10.1103/PhysRevE.110.L023301},
  url = {https://link.aps.org/doi/10.1103/PhysRevE.110.L023301}
}

@article{durve2020learning,
	title = {Learning to flock through reinforcement},
	author = {Durve, Mihir and Peruani, Fernando and Celani, Antonio},
	journal = {Phys. Rev. E},
	volume = {102},
	issue = {1},
	pages = {012601},
	numpages = {8},
	year = {2020},
	month = {Jul},
	publisher = {American Physical Society},
	doi = {10.1103/PhysRevE.102.012601},
	url = {https://link.aps.org/doi/10.1103/PhysRevE.102.012601}
}

@article{peshkov2014boltzmann,
	title={Boltzmann-{G}inzburg-{L}andau approach for continuous descriptions of generic {V}icsek-like models},
	author={Peshkov, Anton and Bertin, Eric and Ginelli, Francesco and Chat{\'e}, Hugues},
	journal={The European Physical Journal Special Topics},
	volume={223},
	number={7},
	pages={1315--1344},
	year={2014},
	publisher={Springer}
}

@article{bertin2020micro,
	title = {Deflection of phototactic microswimmers through obstacle arrays},
	author = {Brun-Cosme-Bruny, Marvin and F\"ortsch, Andre and Zimmermann, Walter and Bertin, Eric and Peyla, Philippe and Rafa\"{\i}, Salima},
	journal = {Phys. Rev. Fluids},
	volume = {5},
	issue = {9},
	pages = {093302},
	numpages = {21},
	year = {2020},
	month = {Sep},
	publisher = {American Physical Society},
	doi = {10.1103/PhysRevFluids.5.093302},
	url = {https://link.aps.org/doi/10.1103/PhysRevFluids.5.093302}
}

@article{nava2018markovian,
  title = {Markovian robots: Minimal navigation strategies for active particles},
  author = {Nava, Luis G\'omez and Gro\ss{}mann, Robert and Peruani, Fernando},
  journal = {Phys. Rev. E},
  volume = {97},
  issue = {4},
  pages = {042604},
  numpages = {17},
  year = {2018},
  month = {Apr},
  publisher = {American Physical Society},
  doi = {10.1103/PhysRevE.97.042604},
  url = {https://link.aps.org/doi/10.1103/PhysRevE.97.042604}
}

@article{martin2016,
	title = {Photofocusing: Light and flow of phototactic microswimmer suspension},
	author = {Martin, Matthieu and Barzyk, Alexandre and Bertin, Eric and Peyla, Philippe and Rafai, Salima},
	journal = {Phys. Rev. E},
	volume = {93},
	issue = {5},
	pages = {051101},
	numpages = {5},
	year = {2016},
	month = {May},
	publisher = {American Physical Society},
	doi = {10.1103/PhysRevE.93.051101},
	url = {https://link.aps.org/doi/10.1103/PhysRevE.93.051101}
}

@article{nasiri2024smart,
  title={Smart active particles learn and transcend bacterial foraging strategies},
  author={Nasiri, Mahdi and Loran, Edwin and Liebchen, Benno},
  journal={Proceedings of the National Academy of Sciences},
  volume={121},
  number={15},
  pages={e2317618121},
  year={2024},
  publisher={National Acad Sciences}
}

@article{jung2025kinetic,
  title = {Kinetic Theory of Decentralized Learning for Smart Active Matter},
  author = {Jung, Gerhard and Ozawa, Misaki and Bertin, Eric},
  journal = {Phys. Rev. Lett.},
  volume = {134},
  issue = {24},
  pages = {248302},
  numpages = {10},
  year = {2025},
  month = {Jun},
  publisher = {American Physical Society},
  doi = {10.1103/5m44-kwhv},
  url = {https://link.aps.org/doi/10.1103/5m44-kwhv}
}

@article{cates2013when,
title = {When are active Brownian particles and run-and-tumble particles equivalent? Consequences for motility-induced phase separation},
author = {M. E. Cates and J. Tailleur},
journal = {Europhys. Lett.},
volume = {101},
pages = {20010},
year = {2013}
}

@incollection{lowen2026towards,
  title={Towards intelligent active particles},
  author={L{\"o}wen, Hartmut and Liebchen, Benno},
  booktitle={Artificial Intelligence and Intelligent Matter: Nanoscience, Soft Matter, Philosophy},
  pages={257--271},
  year={2026},
  publisher={Springer}
}

@article{garnier2025hydrodynamics,
  title={Hydrodynamics of cooperation and self-interest in a two-population occupation model},
  author={Garnier-Brun, J{\'e}r{\^o}me and Zakine, Ruben and Benzaquen, Michael},
  journal={Physical Review Letters},
  volume={135},
  number={10},
  pages={107402},
  year={2025},
  publisher={APS}
}

@article{han2025fluctuation,
  title={Fluctuation theorem and optimal control of an active tracking particle with information processing},
  author={Han, Tai and Meng, Fanlong},
  journal={arXiv preprint arXiv:2508.21487},
  year={2025}
}

@book{te2025artificial,
  title={Artificial intelligence and intelligent matter},
  author={te Vrugt, Michael},
  year={2025},
  publisher={Springer}
}

@article{floyd2024learning,
  title={Learning to control non-equilibrium dynamics using local imperfect gradients},
  author={Floyd, Carlos and Dinner, Aaron R and Vaikuntanathan, Suriyanarayanan},
  journal={arXiv preprint arXiv:2404.03798},
  year={2024}
}

@article{janzen2026active,
  title={Active matter as a framework for living systems-inspired Robophysics},
  author={Janzen, Giulia and Maselli, Gaia and Jimenez, Juan F and Garcia-Perez, Lia and Matoz Fernandez, DA and Valeriani, Chantal},
  journal={Europhysics Letters},
  volume={154},
  number={3},
  pages={37001},
  year={2026},
  publisher={EDP Sciences, IOP Publishing and Societ{\`a} Italiana di Fisica}
}

@article{jung2026theory,
  title={Theory of collective learning in populations of adaptive agents},
  author={Jung, Gerhard and Asnacios, Johann and Ozawa, Misaki and Dauchot, Olivier and Bertin, Eric},
  journal={arXiv preprint arXiv:2607.02171},
  year={2026}
}

@article{mukhopadhyay2026automated,
author = {Aritra K. Mukhopadhyay  and Ran Niu  and Linhui Fu  and Kai Feng  and Christopher Fujta  and Qiang Zhao  and Jinping Qu  and Benno Liebchen },
title = {Automated decision-making by chemical echolocation in active droplets},
journal = {Proceedings of the National Academy of Sciences},
volume = {123},
number = {5},
pages = {e2526773123},
year = {2026},
doi = {10.1073/pnas.2526773123},
URL = {https://www.pnas.org/doi/abs/10.1073/pnas.2526773123},
}

@article{adler1946study,
  title={A study of locking phenomena in oscillators},
  author={Adler, Robert},
  journal={Proceedings of the IRE},
  volume={34},
  number={6},
  pages={351--357},
  year={1946},
  publisher={IEEE}
}

@article{shlomovitz2014phase,
  title={Phase-locked spiking of inner ear hair cells and the driven noisy Adler equation},
  author={Shlomovitz, Roie and Roongthumskul, Yuttana and Ji, Seung and Bozovic, Dolores and Bruinsma, Robijn},
  journal={Interface focus},
  volume={4},
  number={6},
  pages={20140022},
  year={2014}
}

@article{rusch2026intermediate,
  title={Intermediate scattering function of Brownian particles in a tilted cosine potential},
  author={Rusch, Regina and Franosch, Thomas},
  journal={Physical Review E},
  volume={114},
  number={1},
  pages={014127},
  year={2026},
  publisher={APS}
}

@article{de2011modeling,
  title={On the modeling of collective learning dynamics},
  author={De Lillo, Silvana and Bellomo, Nicola},
  journal={Applied mathematics letters},
  volume={24},
  number={11},
  pages={1861--1866},
  year={2011},
  publisher={Elsevier}
}

@article{coscia2011mathematical,
  title={On the mathematical theory of living systems II: The interplay between mathematics and system biology},
  author={Coscia, Vincenzo and Fermo, Luisa and Bellomo, Nicola},
  journal={Computers \& Mathematics with Applications},
  volume={62},
  number={10},
  pages={3902--3911},
  year={2011},
  publisher={Elsevier}
}

@article{burini2016collective,
  title={Collective learning modeling based on the kinetic theory of active particles},
  author={Burini, Diletta and De Lillo, Silvana and Gibelli, Livio},
  journal={Physics of life reviews},
  volume={16},
  pages={123--139},
  year={2016},
  publisher={Elsevier}
}

\end{document}